\documentclass[a4paper,11pt]{article}
\usepackage{pos}
\usepackage{tikz,pgfplots}
\pgfplotsset{compat=newest}
\usepackage[compat=1.1.0]{tikz-feynman}
\selectcolormodel{RGB}
\definecolor{nustblue}{RGB}{0,70,120}
\definecolor{nustblue1}{RGB}{112,110,140} 
\definecolor{nustblue2}{RGB}{190,190,200} 
\definecolor{nustblue3}{RGB}{100,100,255}

\definecolor{bordeaux}{RGB}{123, 0, 44} 
\definecolor{bottlegreen}{RGB}{52, 59, 41}
\definecolor{senape}{RGB}{242, 199, 99}
\definecolor{hawtornrose}{RGB}{136,76,94}
\title{Tuning of QCD+QED simulations with C$^{\star}$ boundary conditions}
\author[a]{Anian Altherr}
\author[c]{Lucius Bushnaq} 
\author[d]{Isabel Campos} 
\author[a]{Marco Catillo} 
\author*[b]{Alessandro Cotellucci}
\author[b,e,f,g]{Madeleine Dale}
\author[c]{Patrick Fritzsch}
\author[a]{Roman Gruber}
\author[b,h]{Jens L\"ucke}
\author[a]{Marina Krstić Marinković}
\author[b,h]{Agostino Patella}
\author[e,f]{Nazario Tantalo}
\author[a]{Paola Tavella}

\affiliation[a]{Institut f\"ur Theoretische Physik, ETH Z\"urich, Wolfgang-Pauli-Str. 27, 8093 Z\"urich, Switzerland}
\affiliation[b]{Humboldt Universit\"at zu Berlin, Institut f\"ur Physik \& IRIS Adlershof, \\Zum Grossen Windkanal 6, 12489 Berlin, Germany}
\affiliation[c]{School of Mathematics, Trinity College Dublin, Dublin 2, Ireland}
\affiliation[d]{Instituto de F\'isica de Cantabria \& IFCA-CSIC, Avda. de Los Castros s/n, 39005 Santander, Spain}
\affiliation[e]{Universit\`a di Roma Tor Vergata, Dipartimento di Fisica, \\Via della Ricerca Scientifica 1, 00133 Rome, Italy}
\affiliation[f]{INFN, Sezione di Tor Vergata, Via della Ricerca Scientifica 1, 00133 Rome, Italy}
\affiliation[g]{University of Cyprus, Department of Physics, 1 Panepistimiou Street, 2109 Aglantzia, Nicosia, Cyprus}
\affiliation[h]{DESY, Platanenallee 6, D-15738 Zeuthen, Germany}

\emailAdd{alecote@physik.hu-berlin.de}

\abstract{We give an update on the ongoing effort of the RC$^\star$ collaboration to generate fully dynamical QCD+QED ensembles with C$^\star$ boundary conditions using the openQ$^\star$D code. The simulations were tuned to the U-symmetric point ($m_d = m_s$) with pions at $m_{\pi^{\pm}} \approx 400$ MeV. The splitting of the light mesons is used as one of three tuning observables and fixed to $m_{K^{0}} - m_{K^{\pm}} \approx 5$ MeV and $m_{K^{0}} - m_{K^{\pm}} \approx 25$ MeV on ensembles with renormalized electromagnetic coupling $\alpha_{\text{R}} \approx \alpha_{\text{phys}}$ and $\alpha_R \approx 5.5\alpha_{phys}$ respectively. The tuning of the three independent quark masses to the desired lines of constant physics is particularly challenging. We will define the chosen hadronic renormalization scheme, and we will present a tuning strategy based on a combination of mass reweighting and linear interpolation to explore the parameter space. We will comment on finite-volume effects comparing meson masses on two different volumes with $m_{\pi^{\pm}} L \approx 3.2$ and $m_{\pi^{\pm}} L \approx 5.1$. We will also provide some technical details on our updated strategy to calculate the sign of the fermionic Pfaffian, which arises in presence of C$^\star$ boundary conditions in place of the standard fermionic determinant. More technical details on the generation of the configurations can be found in J. L\"ucke's proceedings}

\FullConference{%
  The 39th International Symposium on Lattice Field Theory (Lattice2022),\\
  8-13 August, 2022 \\
  Bonn, Germany 
}

\begin{document}
\maketitle

\section{Introduction}
We give an update on the ongoing effort of the RC$^\star$ collaboration to generate fully dynamical QCD+QED ensembles with C$^\star$ boundary conditions~\cite{Kronfeld:1990qu,Kronfeld:1992ae,Wiese:1991ku,Polley:1993bn} and O(a) improved Wilson fermions using the \texttt{openQ$^\star$D} code release $1.1$~\cite{openQxD-csic} at different values of the QED coupling constant. The following discussion is based on the set of ensembles listed in Table \ref{tab:ens}, the technical details on the generation of these configurations can be found in  Jens L\"ucke’s proceedings \cite{JensPOS}.

In this proceedings, we focus on the tuning strategy that we use to generate the configurations, on the mild sign problem of the Pfaffian and how we deal with it in our simulations and on the study of QCD and QED finite-volume effects.
\\
 \begin{table}[h!]
   \centering
      \scalebox{0.8}{\begin{tabular}{ccccccc}
         \hline
         ensemble & lattice &
         $a$ [fm] & $\alpha_R$ &
         Quark Content \\
         \hline
         \hline
         \texttt{A400a00b324} & $64 \times 32^3$ & $0.05393(24)$  & $0$ & $m_u=m_d=m_s\neq m_c$ \\
         \texttt{B400a00b324} & $80 \times 48^3$ & $0.05400(14)$  & $0$ & $m_u=m_d=m_s\neq m_c$ \\
         \hline
         \texttt{A450a07b324} & $64 \times 32^3$ & $0.05469(32)$ & $0.007076(24)$ & $m_u\neq m_d=m_s\neq m_c$ \\
         \texttt{A380a07b324} & $64 \times 32^3$ & $0.05323(28)$ &  $0.007081(19)$ & $m_u\neq m_d=m_s\neq m_c$ \\
         \hline
         \texttt{A500a50b324} & $64 \times 32^3$ & $0.05257(14)$ & $0.040772(85)$ & $m_u\neq m_d=m_s\neq m_c$ \\
         \texttt{A360a50b324} & $64 \times 32^3$ & $0.05054(27)$ & $0.040633(80)$ & $m_u\neq m_d=m_s\neq m_c$ \\
         \texttt{C380a50b324} & $96 \times 48^3$ & $0.050625(79)$ & $0.04073(11)$ & $m_u\neq m_d=m_s\neq m_c$ \\
         \hline
      \end{tabular}}
   \tiny\captionof{table}{List of our ensembles.}
  \label{tab:ens}
   \end{table}

\section{Lines of Constant Physics}
As in pure-QCD, for QCD+QED simulations it is fundamental to define a hadronic renormalization, i.e. a set of observables to keep fixed towards the continuum limit. Our scheme is made of the Wilson-flow-based scale $t_0$, the Wilson-flow fine-structure constant $\alpha_{\text{R}}$ at energy $t_0$ and the dimensionless hadronic quantities:
\begin{equation}
\begin{aligned}
\phi_0=&8t_0\left(M_{K^{\pm}}^2-M_{\pi^{\pm}}^2\right),\\
\phi_1=&8t_0\left(M_{\pi^{\pm}}^2+M_{K^{\pm}}^2+M_{K^{0}}^2\right),\\
\phi_2=&8t_0\left(M_{K^{0}}^2-M_{K^{\pm}}^2 \right)\alpha^{-1}_R,\\
\phi_3=&\sqrt{8t_0}\left(M_{D_{s}}-M_{D^{0}}+M_{D^{\pm}}\right).
\end{aligned}
\label{eq:phi}
\end{equation}
The $\phi$ observables are maximally sensitive to certain combinations of quark masses and, therefore, they are useful for the tuning of the bare quark masses.

\begin{figure}[h!]
\centering      
      \includegraphics[scale=0.6]{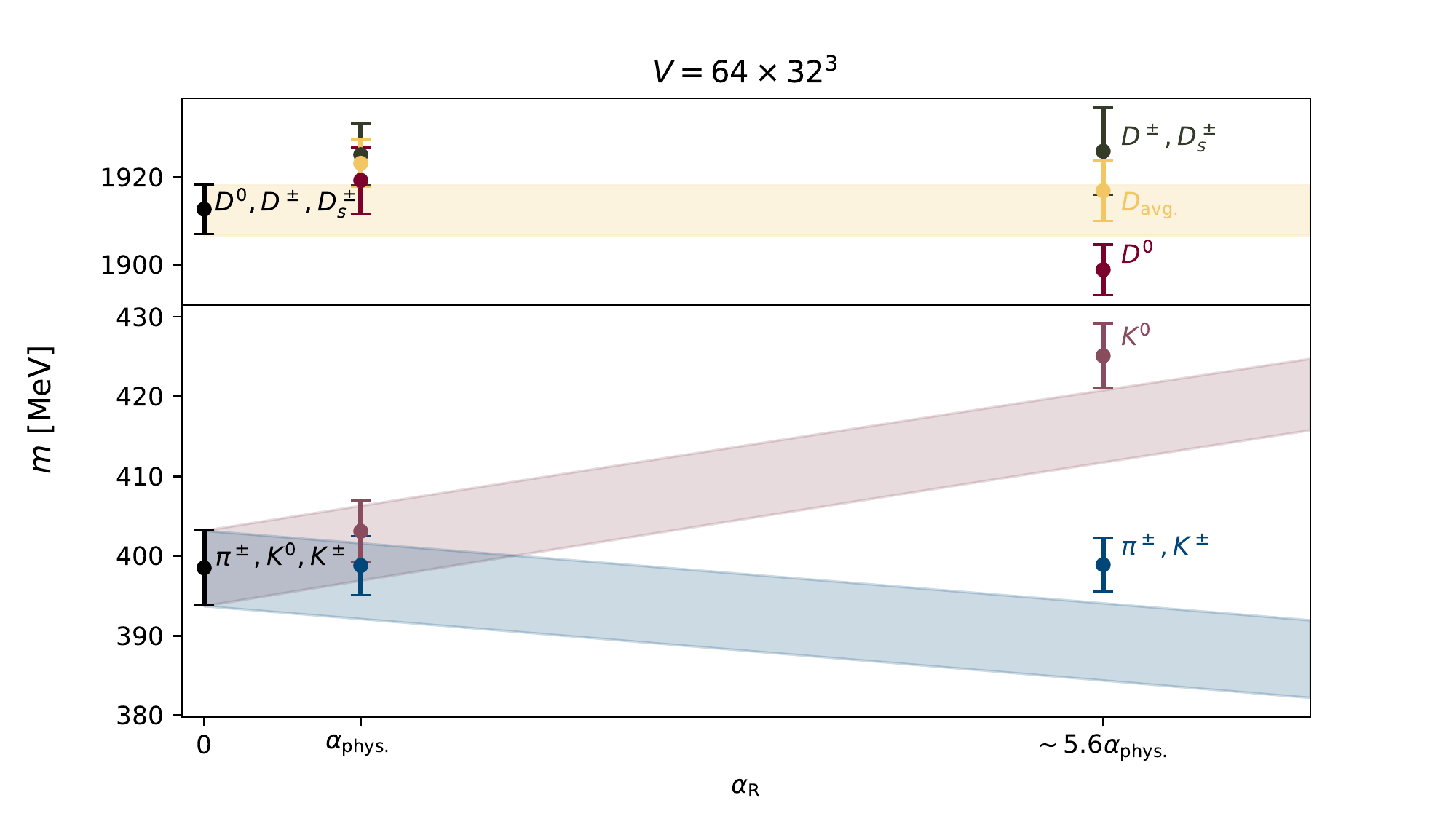}
      \captionof{figure}{Lines of constant physics.}
    \label{fig:lines}
\end{figure}

The scale is set using the central value measured in QCD $\left(8t_0\right)^{1/2}=0.415$ fm by CLS \cite{Bruno:2016plf}. This value is not the most accurate determination in QCD, see e.g. \cite{Strassberger:2021tsu}, and it is calculated neglecting isospin corrections. Eventually, we will need to set the scale consistently in QCD+QED, e.g. by means of some baryon mass.
Substituting the masses from the PDG \cite{ParticleDataGroup:2020ssz} and the reference value for $t_0$ in Eq. \eqref{eq:phi}, the values of the physical $\phi$s are:
\begin{equation}
\phi_0^{\text{phys}}\simeq 0.992,\ \phi_1^{\text{phys}}\simeq 2.26,\ \phi_2^{\text{phys}}\simeq 2.36,\ \phi_3^{\text{phys}}= 12.0.
\end{equation}
We simulate as close as possible to these values, up to $\phi_0$ which is set to zero to take advantage of the U-spin symmetric point where $m_d=m_s$. The values used to generate the configurations are:
\begin{equation}
\phi_0=0,\ \phi_1= 2.11,\ \phi_2= 2.36,\ \phi_3= 12.1.
\end{equation}
Keeping them constant while varying the value of $\alpha_{\text{R}}$ we define our \textit{lines of constant physics} plotted in Fig. \ref{fig:lines}.

\section{Tuning Strategy}
For fixed values of the bare fine-structure constant $\alpha$ and of the inverse SU(3)-gauge coupling $\beta$, we tune the bare quark masses to obtain the desired values of the $\phi$ observables. Since we work with $m_d = m_s\equiv m_{ds}$, this is a three-parameter tuning problem (which turned out to be expensive and challenging). 
\\
\begin{figure}[h!]
\tikzstyle{block} = [draw, rectangle, minimum height=3em, minimum width=3em]
\tikzstyle{virtual} = [coordinate]
\centering
\scalebox{0.7}{
\begin{tikzpicture}[auto, node distance=2.6cm]
    \node [virtual]                 (input)     {};
    \node [draw, right=0.7cm of input,minimum width=2cm, minimum height=1.3cm, align=center] (IntEns)    {Generate\\Ensemble\\(1000 cnfg)};
    
    \node [block, minimum width=2.3cm, minimum height=1.3cm,right of=IntEns, align=center] (RW)    {3 Reweighting\\one for each\\$\hat{m}_i+\Delta m_i$};
    
    \node [virtual, above=0.7cm of RW] (deltaM)    {};
    \node [block, right of=RW,minimum width=2.3cm, minimum height=1.3cm, align=center] (Int)    {Interpolation\\$\phi^T = A \tilde{m} + b$};
    \node [block, right=0.7cm of Int,minimum width=2.3cm, minimum height=1.3cm, align=center] (FullStat)    {Generation\\full statistics\\(2000 cnfg)};
    \node [draw,diamond,aspect=2, below of=FullStat, align=center] (Condition)    {$\phi_i=\phi^T$?};
    \node [draw=none, fill=none, right=1cm of Condition] (End)  {Tuned};
    \draw [-stealth] (input) -- node {$\hat{m}_i$}(IntEns);
    \draw [-stealth] (deltaM) -- node {$\Delta m_i$} (RW);
    \draw [-stealth] (IntEns) -- node {} (RW);
    \draw [-stealth] (RW) -- node {} (Int);
    \draw [-stealth] (Int) -- node {$\tilde{m}_i$} (FullStat);
   \draw [-stealth] (FullStat) -- node {$\phi_i$} (Condition);
   \draw [-stealth] (Condition) -| node[near start,above=0.1cm] {NO} (RW);
    \draw [-stealth] (Condition) -- node[pos=0.4,above=0.1] {YES} (End);
\end{tikzpicture}
}
\captionof{figure}{Conceptual flow chart of the tuning strategy.}
  \label{fig:Tuning}
\end{figure}
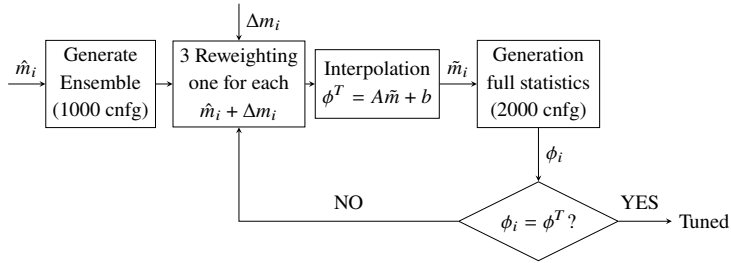
\\
Our tuning strategy takes into account the contribution of the sea quarks by means of reweighting factors in the masses~\cite{RC:2021tah}. The main idea is to start from a guess of the bare masses $\hat{m}_i$ with $i=u,ds,c$ and use them to generate a limited statistics ensemble ($1000$ configurations). Then, with a guess for the shift $\Delta m_i$, we measure one reweighting factor for each mass shift. On each of these four points in the bare parameter space (the original point plus the three shifted ones), we measure the $\phi$ observables and we interpolate or extrapolate to the tuned point to obtain the bare masses $\tilde{m}_i$. In the end, we generate an ensemble at full statistics ($2000$ configurations) using $\tilde{m}_i$ and we check if the masses are correctly tuned, if not we measure other reweighting factors. Fig. \ref{fig:Tuning} shows the conceptual steps of this strategy.

\section{Sign of the Pfaffian}
Instead of the usual determinant of the Dirac operator, as a consequence of C$^{\star}$ boundary conditions, we have to deal with the fermionic Pfaffian. As it is in QCD for the determinant of the Dirac operator with Wilson fermions, the sign of the Pfaffian can flip. This is a mild sign problem, which disappears in the continuum limit. In our simulations, the absolute value of the Pfaffian is included in the RHMC while the sign is incorporated in the analysis as a reweighting factor. To compute the sign, the Pfaffian can be written as the product of the eigenvalues $\lambda_n$s of the hermitian Dirac operator $Q=\gamma_5D$. Due to C$^{\star}$ boundary conditions $Q$ is a $24V\times24V$ matrix but the eigenvalues are doubly degenerate thus we pick each one only once  (see \cite{Bushnaq:2022aam}):
\begin{equation}
    \text{pf}(CKD)=\prod_{n=1}^{12V}\lambda_n
\end{equation}
where $CKD$ is the Dirac operator with C$^{\star}$ boundary conditions. For a high value of the mass $M$, $Q\sim M\gamma_5$ hence the negative eigenvalues are even so the determinant is positive. In fact, the sign of the Pfaffian is estimated following the $\lambda_n$s flow from the target mass $m$ to some arbitrarily larger mass $M$. One can see that there is a relationship between the number of times $\lambda_n$ crosses zero and the final sign of the Pfaffian
\begin{equation}
\begin{aligned}
&\text{number of }\lambda_n\ \text{crossing zero even}\rightarrow\text{positive sign} \\   
&\text{number of }\lambda_n\ \text{crossing zero odd}\rightarrow\text{negative sign}.\notag 
\end{aligned}
\end{equation}
\begin{figure}[h!]
\centering
      \includegraphics[scale=0.48]{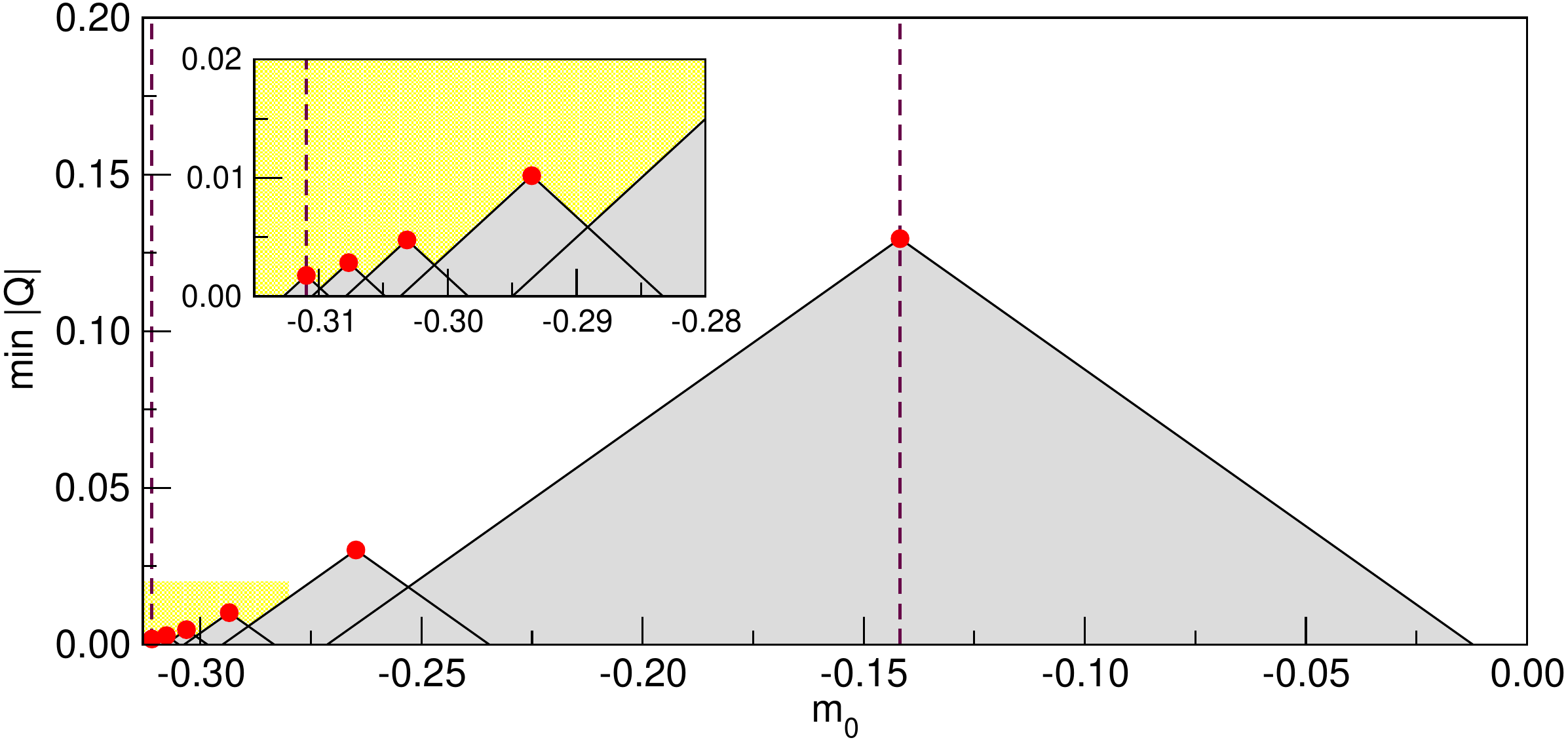}
      \captionof{figure}{Smallest eigenvalue of up quark $|Q|$ as a function of the valence mass, calculated on \texttt{C380a50b324}.}
  \label{fig:scanpf}
\end{figure}
\\
As shown in Fig. \ref{fig:scanpf}, we follow only the flow of the smallest eigenvalue of $|Q|$ for different values of the valence mass. This is sufficient because no eigenvalue can flow in the grey regions of (Fig. \ref{fig:scanpf}) (as demonstrated in \cite{Bushnaq:2022aam}). If the smallest eigenvalue increases the grey region keeps growing and covers all the values of the masses where no flip can occur. The flow can be also followed starting from the largest considered mass $M$. We follow the flow from both sides and we choose to continue with the side which has the biggest step in the mass. At every step, the interval in which a sign flip may occur shrinks. If the smallest eigenvalue shows non-monotonicity with the mass, we stop this algorithm and track the eigenvalues and eigenvectors flow in the restricted interval, using the methods described in \cite{Campos:1999du,Mohler:2020txx} (for a detailed description of the algorithm see \cite{Bushnaq:2022aam}).

\begin{figure}[h!]
  \setlength{\tabcolsep}{-0.2em}
\centering
  \begin{tabular}[c]{cc}
  \includegraphics[scale=0.39]{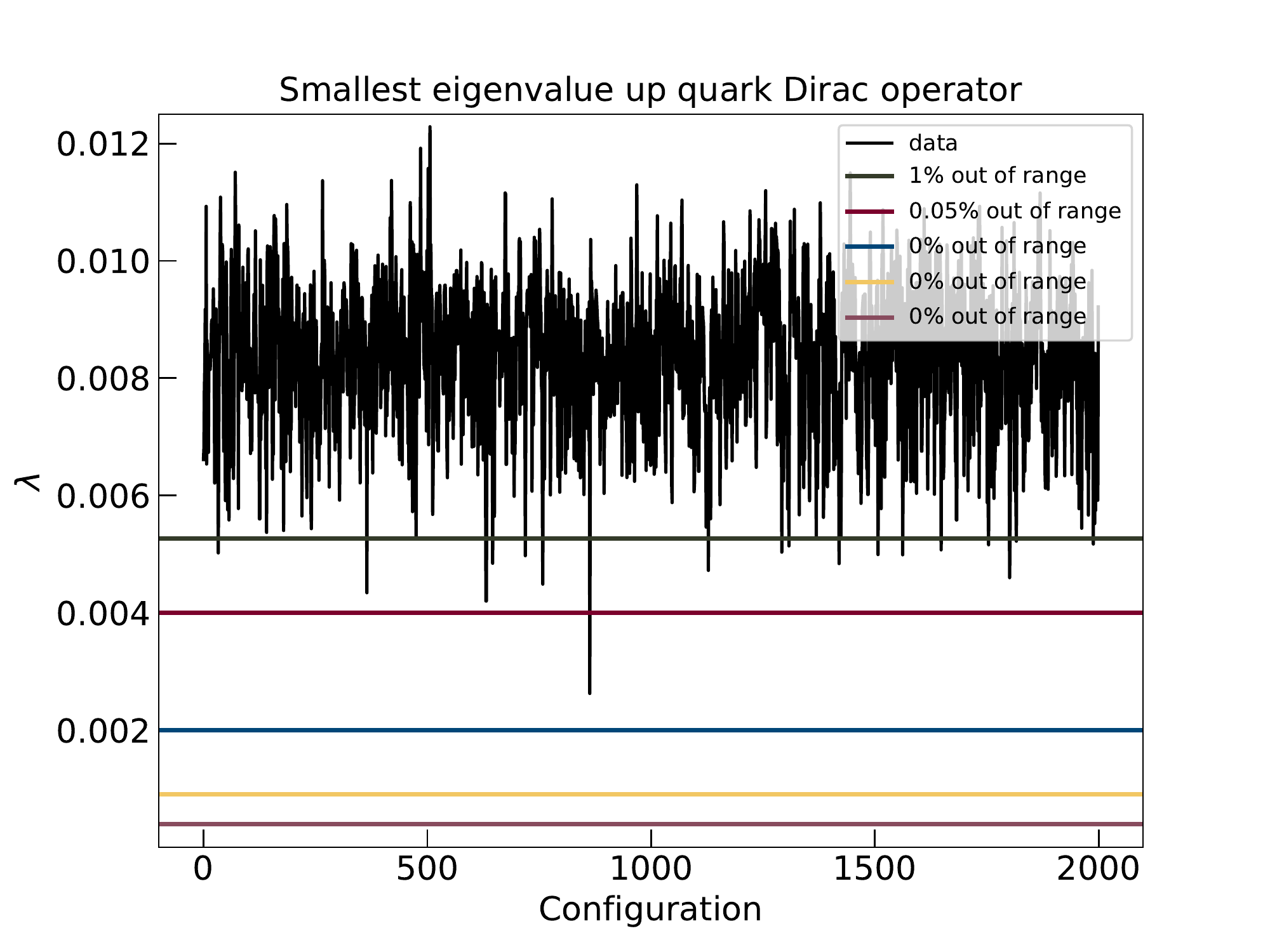}&
  \includegraphics[scale=0.49]{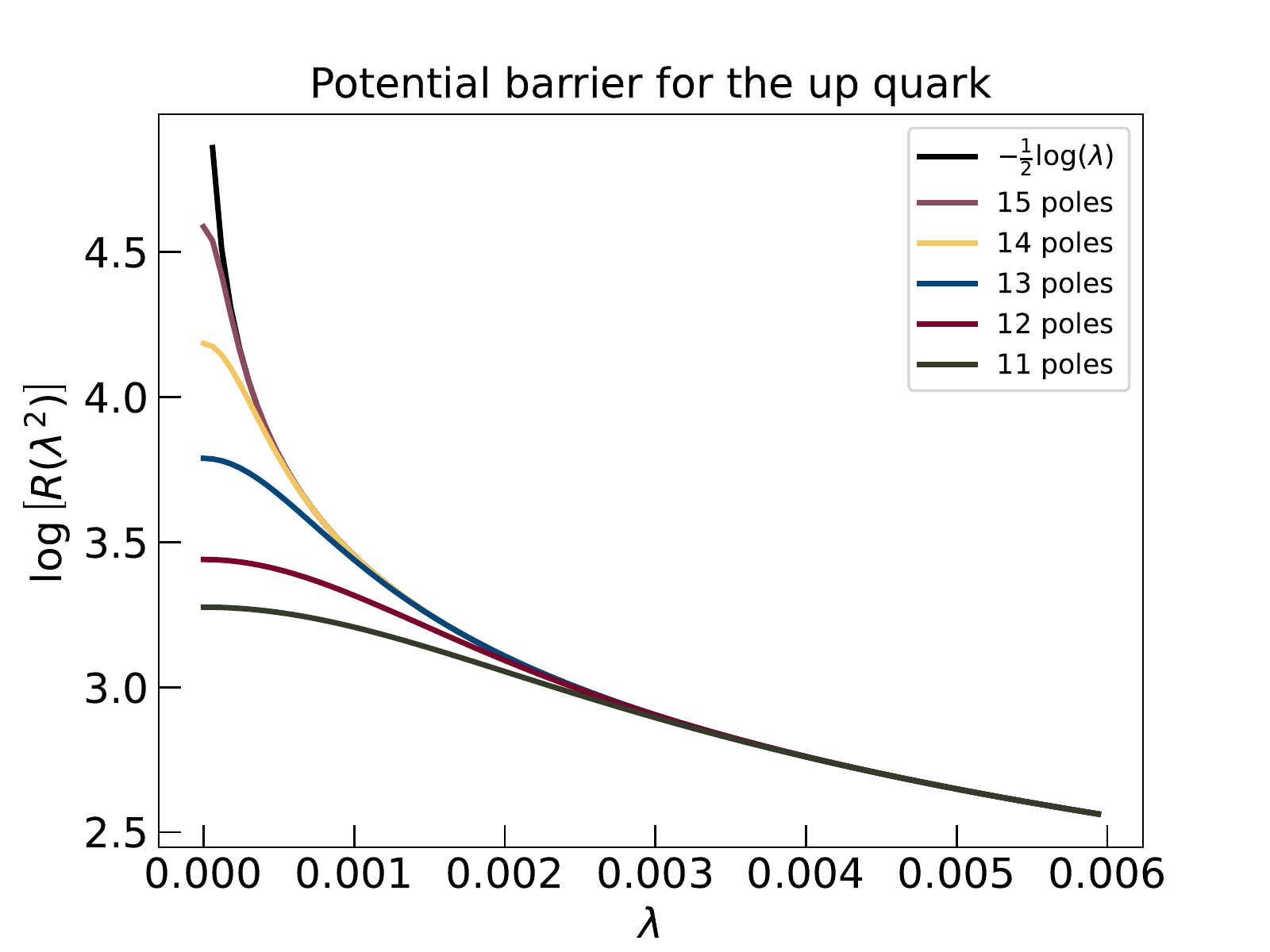}\\
  \end{tabular}
  \captionof{figure}{Smallest eigenvalue of the up quark Dirac operator of \texttt{A450a07b324} with different possible rational approximations (left). The potential barrier for the sign flip of the up quark Pfaffian for different rational approximations (right). In (right) also the logarithm is plotted which is the behaviour of the barrier without the rational approximation.}
  \label{fig:figlabel}
\end{figure}

In our case, we compute only the sign of the up quark Pfaffian because the down and strange quarks are degenerate hence they contribute with a squared Pfaffian which is positive, and the charm quark is massive enough not to show sign flips. The probability of the sign flip depends on the potential barrier in Fig. \ref{fig:figlabel} which is determined by the rational approximation chosen for the Dirac operator \cite{Mohler:2020txx}. The rational approximation is defined by a range of validity and some poles, the barrier can be lowered by increasing the lower bound of validity, which requires reducing the number of poles.

We generated \texttt{A450a07b324} with a 15-poles rational approximation (purple line in Fig. \ref{fig:figlabel}) and 
we saw no sign flip so we decided to lower the barrier to 13-poles (blue line in Fig. \ref{fig:figlabel}) to generate \texttt{A380a07b324}. The result is still no sign flip, the reason is under study but we propose three possible sources. The generation algorithm is stuck in a region of the configuration space where there is no flip so it may be not ergodic. The lattice spacing is small enough to have a small artefact due to Wilson fermion. The up quark is too massive to have a negative eigenvalue, however this is unlikely, since CLS has already shown that a strange quark as massive as our up quark can give rise to sign flips \cite{Mohler:2020txx}.

\section{Finite-Volume Effects}
\begin{figure}[h!]
\centering
      \includegraphics[scale=0.55]{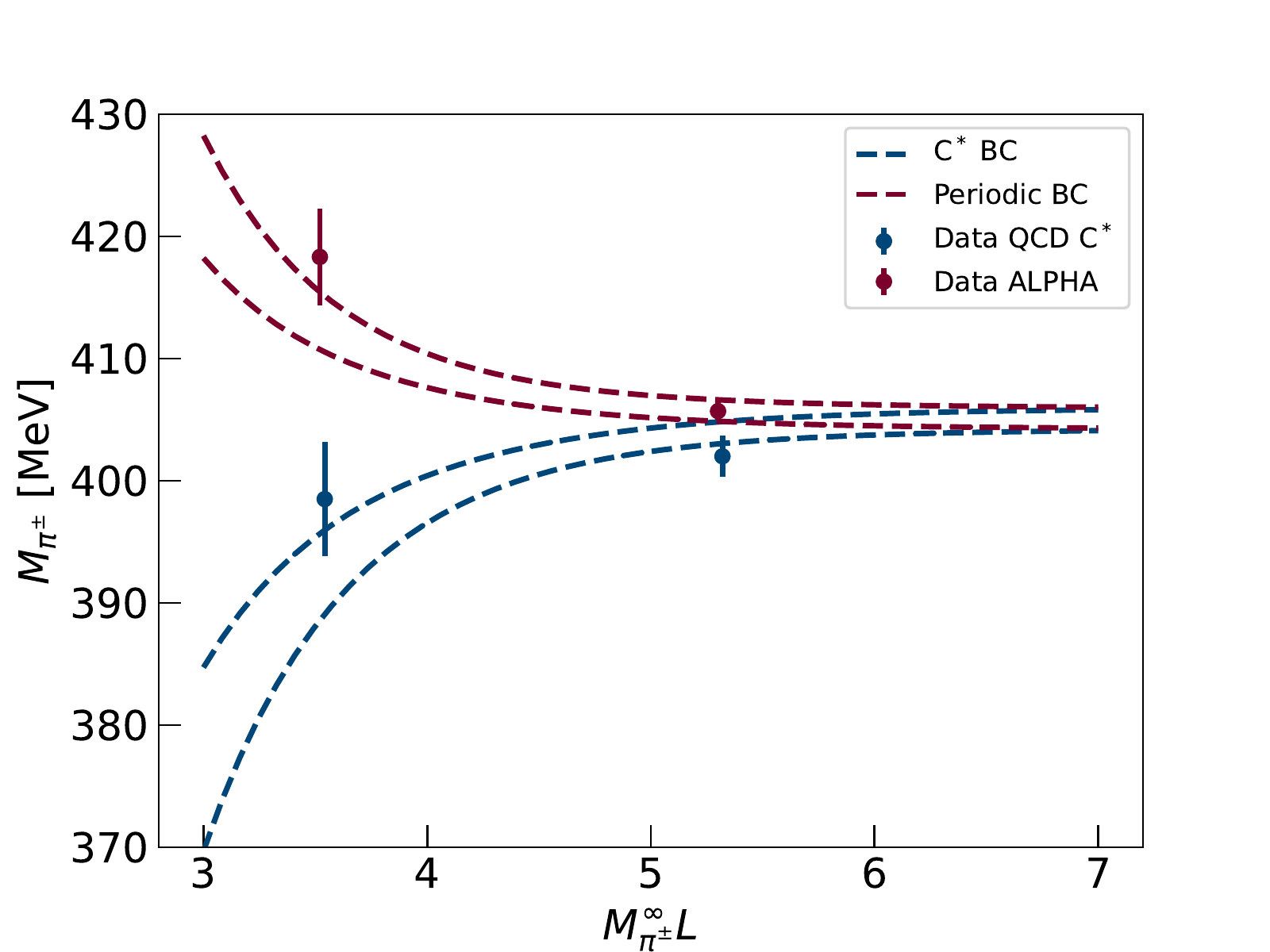}
      \captionof{figure}{Pion masses for our QCD
      ensembles, and ALPHA ensembles from~\cite{Hollwieser:2020qri}.}
        \label{fig:FVEQCD}
\end{figure}

QCD finite-volume effects are exponentially suppressed for stable hadron masses \cite{Luscher:1986pf}, we estimated them by means of LO $\chi$-PT:
\begin{equation}
\begin{aligned}
    \color{bordeaux} M_{\text{P}}(L) =&\color{bordeaux} M + \frac{\xi}{3} \sum_{\vec{n} \in \mathbb{Z}^3 \setminus \{0\}} \frac{2}{nL} K_1(nL)
   \ ,\notag \\
    \color{nustblue} M_{\text{C}}(L) =&\color{nustblue} M - \frac{\xi}{3} \sum_{\vec{n} \in \mathbb{Z}^3 \setminus \{0\}} \frac{1 - 3(-1)^{\sum_i n_i}}{nL} K_1(nL)
   \ ,\notag
\end{aligned}
\end{equation}
where $n=|\vec{n}|$, $\xi = M^2/(4 \pi F)^2$ and $K_1$ is a modified Bessel function of the second kind, for \textcolor{bordeaux}{periodic}~\cite{GASSER1987477} and \textcolor{nustblue}{C$^\star$} boundary conditions. On the ensemble \texttt{B400a00b324}, the QCD finite-volume effects on the pion mass are smaller than $1\%$ (so negligible for our target precision). As one can see in Fig. \ref{fig:FVEQCD}, the pion mass of the ensemble \texttt{A400a00b324} is probably strongly affected by finite-volume
effects.

\begin{figure}[h!]
\centering
  \setlength{\tabcolsep}{-0.5em}
  \begin{tabular}[c]{cc}
  \includegraphics[scale=0.4]{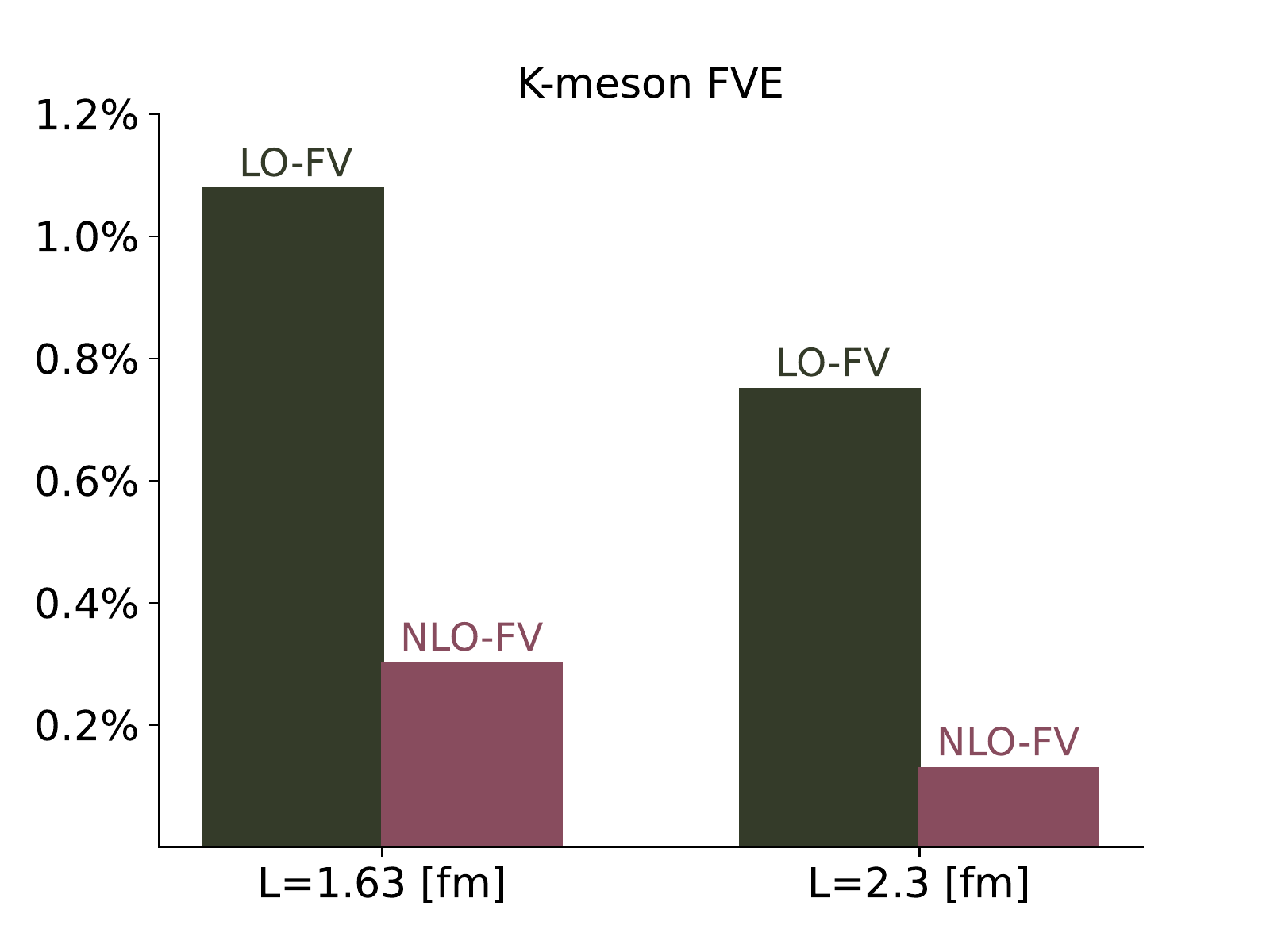}&
  \includegraphics[scale=0.4]{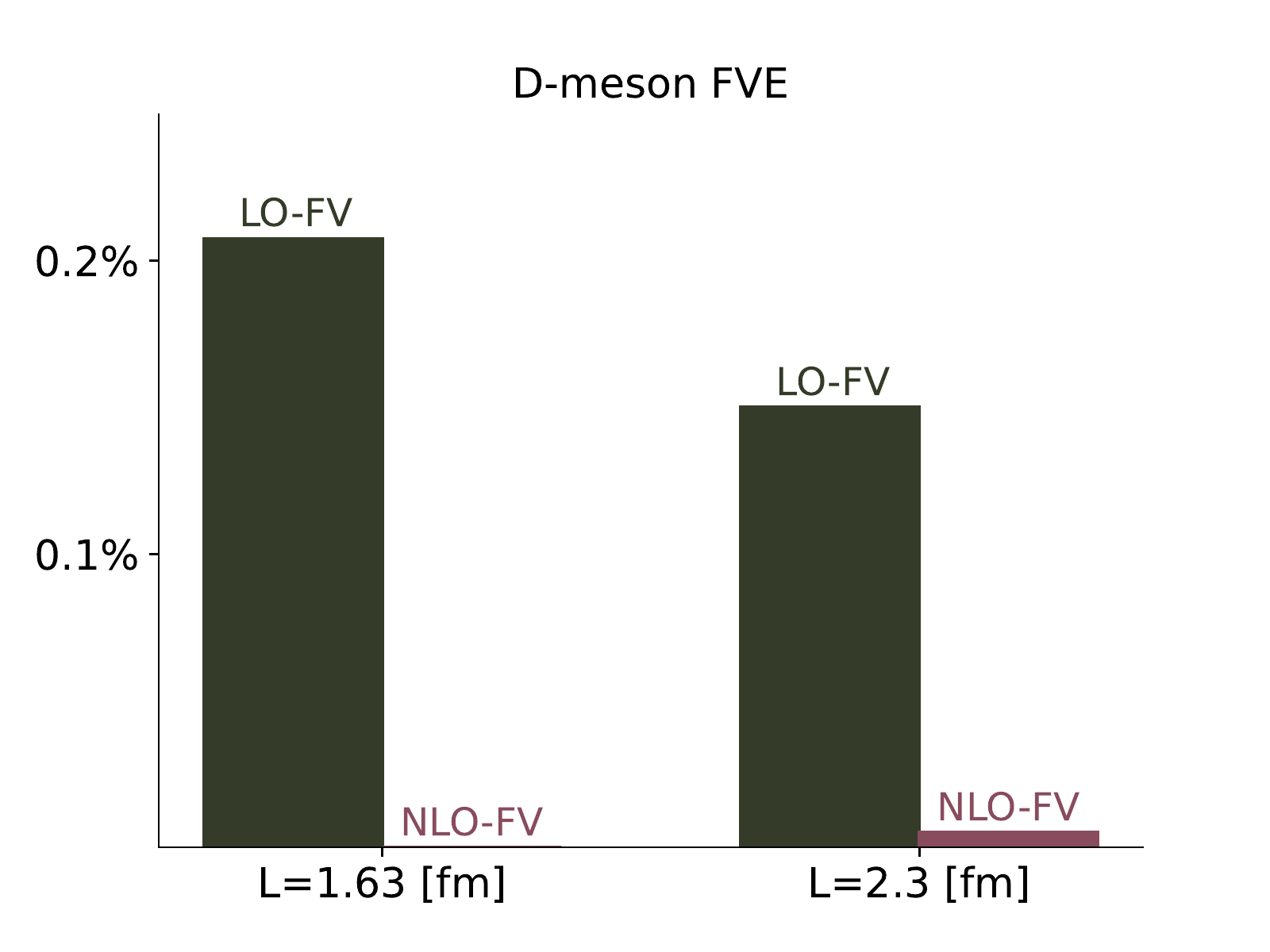}\\
  \end{tabular}
  \captionof{figure}{QED finite-volume effects for the chartged meson masses for $\alpha_R\approx5.5\alpha_{\text{phys}}$.}
  \label{fig:FVEQED}
\end{figure}

QED finite-volume effects for hadron masses have a power-law behaviour as derived in~\cite{Lucini:2015hfa}:
\begin{equation}
   M_{\text{C}}(L) = M - \alpha \left\{
   \color{bottlegreen}\frac{q^2 \zeta(1)}{2 L}\color{black} + \color{hawtornrose}\frac{q^2 \zeta(2)}{\pi M L^2}\color{black}
   + o\left(\frac{1}{L^4}\right)\right\}\notag ,
\end{equation}
where $\zeta(x)$ is a particular definition of the zeta function described in \cite{Lucini:2015hfa}. The leading order and the next to leading order terms are structure-independent and are subtracted from the final result of the analysis, while the higher-order terms in $1/L$ depend on the derivatives of the Compton tensor. A detailed comparison of LO and NLO finite-volume effects is shown for charged K and D mesons in Fig. \ref{fig:FVEQED}.

As one can see from Fig. \ref{fig:FVEQED} the NLO finite-volume effects are already below the percentage level for this unphysically big value of the QED coupling constant.

\section{Conclusion and outlook}
We showed some of the challenges we faced during the generation of QCD+QED ensembles at the physical value of $\alpha_{\text{EM}}$ that will be important for future works, in particular, the tuning strategy that we designed for the three parameters problem that we also plan to apply to the four non-degenerate quarks case.

Furthermore, from the hierarchy of the finite-volume
effects as shown in Fig. \ref{fig:FVEQED}, we suppose that the contributions of the non-universal terms in the 1/L expansion are in the sub-$0.1\%$ range for values of $m_{\pi^{\pm}}L\sim 5$.

For the sign of the Pfaffian, we described the algorithm that allows us to compute the sign for every configuration generated.

The next step will be to split the down and the strange quark and to go closer to the physical pion mass. We are also implementing the perturbative evaluation of isospin-breaking effects for a cost comparison.

\acknowledgments 
AC and JL research is funded by the Deutsche Forschungsgemeinschaft (DFG, German Research Foundation) - Projektnummer 417533893/GRK2575 ``Rethinking Quantum Field Theory''. The funding from the European Union’s Horizon 2020 research and innovation program under grant agreement No. 813942 and the financial support by SNSF (Project No. 200021\_200866) is gratefully acknowledged.
The authors gratefully acknowledge the computing time granted by the Resource Allocation Board and provided on the supercomputer Lise and Emmy at NHR@ZIB and NHR@Göttingen as part of the NHR infrastructure. The calculations for this research were partly conducted with computing resources under the project bep00085 and bep00102. The work was supported by CINECA that granted computing resources on the Marconi supercomputer to the LQCD123 INFN theoretical initiative under the CINECA-INFN agreement. The authors acknowledge access to Piz Daint at the Swiss National Supercomputing Centre, Switzerland under the ETHZ's share with the project IDs go22, go24, eth8, and s1101. The work was supported by the Poznan Supercomputing and Networking Center (PSNC) through grant numbers 450 and 466.

\bibliographystyle{JHEP}
\bibliography{inspire}

\end{document}